\documentclass[aps,pre,reprint]{revtex4-2}
\usepackage{amsmath}
\usepackage{amssymb}
\usepackage{graphicx}

\newcommand{\e}{\varepsilon}              %
\newcommand{\w}{{\omega}}

\begin{document}


\title{Conservative dynamics in phase oscillator networks}

\author{Arkady Pikovsky}
\affiliation{Department of Physics and Astronomy, University of Potsdam, Karl-Liebknecht-Str. 24/25, 14476, Potsdam-Golm, Germany}

\email{pikovsky@uni-potsdam.de}

\begin{abstract}
The interaction between phase oscillators is conservative if the phase volume is conserved throughout the dynamics. We derive a general condition, based on the notion of a pair-Hamiltonian, for the pairwise couplings to be conservative. The conservative networks with Winfree-type and Kuramoto-Daido-type couplings are also discussed. It is demonstrated that although, in contradistinction to genuine Hamiltonian dynamics, there is no exact pairwise symmetry of the Lyapunov exponents, the Lyapunov spectrum for a large network is nearly symmetric. The concept is also generalized to triplet and quadruplet couplings. 
\end{abstract}
\maketitle
\section{Introduction}
\label{sec:intro}
The interaction between self-sustained autonomous oscillators is often reduced to phase dynamics. With respect to the phases, one speaks about attractive coupling if two oscillators synchronize in a state with close phases (in-phase synchrony), and about repulsive coupling if the phases of synchronous oscillators differ by $\approx \pi$ (out-of-phase synchrony). Such a distinction appears rather arbitrary for two units, because phases can be redefined with an arbitrary constant shift. However, for large populations, distinguishing between attractive and repulsive coupling is meaningful because only the former leads to global synchronization~\cite{Pikovsky-Rosenblum-Kurths-01}. The nature of the coupling can be controlled in experiments, as it, e.g., was done for laser arrays in \cite{Nixon_etal-13}. In the literature, the setups with a mixture of two types of coupling have also been considered~\cite{Hong-Strogatz-11,hong2012mean}. 

Between the two types, one has a neutral/reactive/dispersive coupling, which is neither attractive nor repulsive. Such a coupling between nearest neighbors on a lattice leads to propagating waves (compactons) as has been discussed in \cite{Rosenau-Pikovsky-05,Pikovsky-Rosenau-06,Ahnert-Pikovsky-08,Lee-Cross-11,Rosenau-Pikovsky-20}. In particular, Ref.~\cite{Pikovsky-Rosenau-06} shows that the lattice phase equations can be formulated as a Hamiltonian system. The Hamiltonian-type dynamics has also been reported for a model with star-type coupling~\cite{burylko2021collective}.

The goal of this paper is to provide a general framework for the description of neutral coupling in phase oscillator networks. We define conservative interaction as one that obeys conservation of the phase volume; sometimes this property is referred to as Liouvillian one (cf.~\cite{ashwin2016identical}, where phase volume conservation is discussed for a Kuramoto-type model). This property is weaker than the existence of a Hamiltonian, but it is strong enough to exclude the existence of attractors and repellers. In this respect, it is stronger than reversibility (discussed for phase oscillators in  \cite{tsang1991reversibility,Topaj-Pikovsky-02,altmann2006nontwist,%
ashwin2016identical,burylko2018coexistence,burylko2023time}), where attractors and repellers can coexist with Hamiltonian-like dynamics. 

Consideration of conservative (Liouvillian) systems outside of the Hamiltonian formalism has a long tradition in physics. Probably, the best-known example is Lagrangean particle motion in a three-dimensional incompressible fluid flow. Nambu in 1973~\cite{Nambu-73} introduced a generalization of the Hamilton dynamics to systems of three variables, using two Hamiltonian functions (for recent
developments of this formalism see~\cite{sardon2025quasi}). Nowadays, one witnesses a revival of interest in conservative nonlinear systems in the realm of nonlinear electronic circuits~\cite{Leng-22,Dong-23,Yan-24,minati25three}.  The reason is that chaos in such systems is rather robust, because periodic attractors are impossible; as a consequence, conservative chaotic electronic circuits appear to be optimal for chaos-based cryptography, pseudo-random number generation, and secure communications~\cite{Leng-22,minati25three}.

The goal of this paper is to introduce a novel, rather general framework for the description of conservative phase dynamics of coupled oscillators. The equations of the phase dynamics are rather specific and cannot be described, e.g., within the Nambu formalism. For a system of $N$ oscillators, we introduce generally different coupling pair-Hamiltonians for each interaction (so there are altogether $N(N-1)/2$ pair-Hamiltonians) and derive the phase equations from these functions (plus natural frequencies). We discuss in detail how the conservative phase-dynamics equations look for Winfree-type and Kuramoto-type couplings. As it is expected for general conservative systems of high enough dimension, conservative phase oscillators demonstrate robust chaos, which we characterize by virtue of the spectrum of Lyapunov exponents. Furthermore, we show how the concept can be generalized to many-body interactions, and give simple examples of conservative dynamics with triplet and quadruplet couplings. 

\section{Two oscillators with conservative coupling}
Suppose we have two phase oscillators (variables $x,y$) with natural frequencies $\w_x,\w_y$. The interaction of these
oscillators is described by a system
\begin{equation}
\begin{aligned}
\dot x&=\w_x+f(x,y)\;,\\
\dot y&=\w_y+g(y,x)\;,
\end{aligned}
\label{eq:1}
\end{equation}
where $f,g$ are coupling functions, $2\pi$-periodic in both arguments. Without loss of generality, we assume that coupling functions $f,g$ do not contain constant terms, because those can be absorbed in the natural frequencies.

We now demand that the phase volume is conserved in the course of the dynamics:
\begin{equation}
\partial_x\dot x+\partial_y \dot y=\partial_x f(x,y)+\partial_y g(x,y)=0\;.
\label{eq:2}
\end{equation}
From this condition, it follows that the coupling functions are not independent, but can be generally represented as derivatives of a single function
\begin{equation}
f(x,y)=\partial_y h(x,y),\qquad g(x,y)=-\partial_x h(x,y)\;.
\label{eq:3}
\end{equation}
The resulting equations for coupled oscillators 
\begin{equation}
\begin{aligned}
\dot x&=\w_x+\partial_y h(x,y)\;,\\
\dot y&=\w_y-\partial_x h(x,y)\;,
\end{aligned}
\label{eq:4}
\end{equation}
resemble a Hamiltonian system. However, only the coupling terms are represented as derivatives (formally, one could also represent the frequencies as derivatives by introducing $\tilde h(x,y)=\w_x y-\w_y x+h(x,y)$, but this function is not $2\pi$-periodic in its arguments and thus we prefer to separate natural frequencies and coupling in \eqref{eq:4}). Due to this analogy, and in view of generalizations to networks below, 
we call the coupling function $h(x,y)$ ``pair-Hamiltonian''. 

Equation \eqref{eq:4} is the most general form of the conservative coupling of two phase oscillators, but in the theory of coupled oscillators, coupling functions with a particular structure are often considered, namely Winfree-type and Kuramoto-Daido-type couplings.

\subsection{Winfree-type conservative coupling}
In the Winfree-type coupling~\cite{Winfree-80}, the coupling functions $f,g$ are products of functions of $x$ and $y$. Correspondingly, we assume that the coupling pair-Hamiltonian is represented as
\begin{equation}
h(x,y)=A(x)B(y)\;,
\label{eq:5}
\end{equation}
where $A(x)$ and $B(y)$ are $2\pi$-periodic functions. The conservative dynamics of coupled oscillators now reads
\begin{equation}
\begin{aligned}
\dot x&=\w_x+A(x)B'(y)\;,\\
\dot y&=\w_y-B(y)A'(x)\;.
\end{aligned}
\label{eq:6}
\end{equation}

\subsection{Kuramoto-Daido-type conservative coupling}
In the Kuramoto-Daido-type coupling~\cite{Kuramoto-75,daido1992order}, the coupling functions depend only on the differences of the phases. Accordingly, we set $h(x,y)=H(x-y)$ and obtain
\begin{equation}
\begin{aligned}
\dot x&=\w_x-F(x-y)=\w_x-F_{odd}(x-y)-F_{even}(x-y)\\
\dot y&=\w_y-F(x-y)=\w_y+F_{odd}(y-x)-F_{even}(y-x)
\end{aligned}
\label{eq:7}
\end{equation}
where $F(x)=H'(x)$. Here, we also wrote the equations by decomposing the coupling function into odd and even parts with respect to the phase difference.

\subsection{Illustration of the conservative dynamics}
We note here that in the usual approaches to deriving the phase equations from the original coupled oscillators' equations, one assumes that the coupling is small. In the context of Eqs.~\eqref{eq:1}, this means smallness of terms $f,g$ in comparison to the natural frequencies. However, formally, the considerations above are not restricted to such a smallness; furthermore, one can formally consider equations~\eqref{eq:1} with small, vanishing, or negative natural frequencies. Therefore, in the numerical illustrations below, we will not restrict ourselves to the case of weak coupling.

\begin{figure}
\centering
\includegraphics[width=\columnwidth]{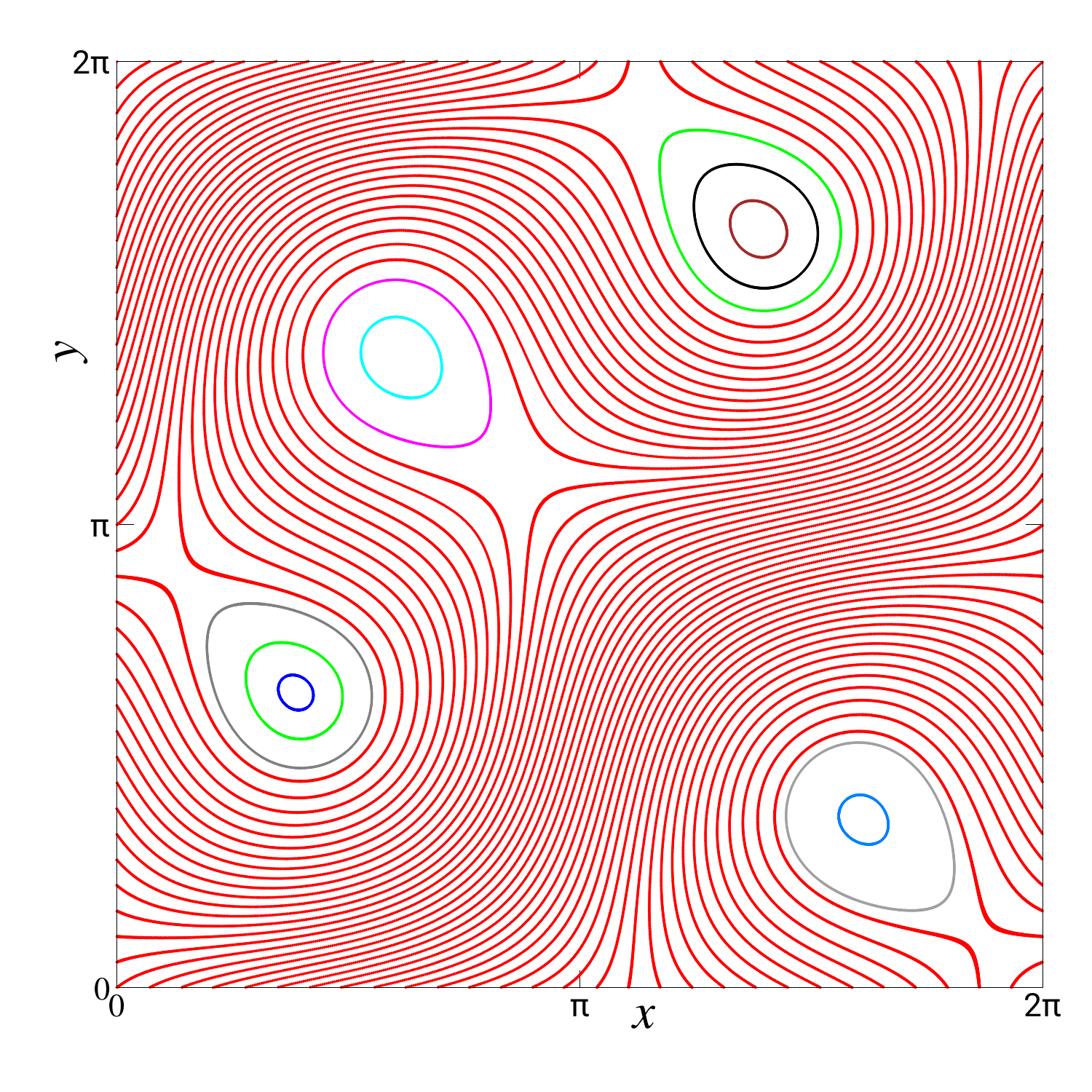}
\caption{Trajectories at the conservative Winfree-type coupling of two oscillators. Parameters are given in text.}
\label{fig:1}
\end{figure}

Conservative equations on a two-dimensional torus \eqref{eq:4} generally produce, depending on initial conditions, quasiperiodic and periodic trajectories, or steady states of a saddle and a center type. We illustrate this in Fig.~\ref{fig:1} for the Winfree-type coupling \eqref{eq:6} with $A(x)=B(x)=\sin(x)$ and natural frequencies $\w_x=0.39,\;\w_y=\pi/10$.

\section{Oscillator networks with conservative pairwise coupling}
In this section, we extend the notion of the conservative interaction to networks with pairwise (two-body) couplings. 

\subsection{General formulation}

We will assume that equations of type \eqref{eq:4} are valid for each interaction. This means that for a network of $N$ oscillators (phases $x_1,.\ldots,x_N$), there are $N(N-1)/2$ pair-Hamiltonians $h_{kj}(x_k,x_j)$ (where $j>k$) and the dynamics of the phases reads
\begin{equation}
\dot{x}_k=\w_k+\sum_{j>k} \partial_j h_{kj}(x_k,x_j)-\sum_{j<k}\partial_j h_{jk}(x_j,x_k)\;.
\label{eq:8}
\end{equation}

In the expression \eqref{eq:8} all couplings are generally different, but it is instructive to look at the cases where all the couplings are similar. In the case of Winfree-type coupling, the simplest form of pair-Hamiltonians is the case where both multipliers are the same function: $h_{kj}=a_{kj}A(x_k)A(x_j)$. Then the network equations have the form
\begin{equation}
\dot{x}_k=\w_k+A(x_k)\sum_{j\neq k} b_{kj} A'(x_j),\quad b_{kj}=\begin{cases}
a_{kj} & j>k\;,\\ -a_{jk} & j<k\;,
\end{cases}
\label{eq:9}
\end{equation}
and $b_{kj}=-b_{jk}$ is an arbitrary real antisymmetric matrix.
The anti-symmetry of the coupling matrix is essential for the coupling to be conservative. Indeed, in a recent Ref.~\cite{li2025associative}, a variant of a Winfree-type coupled oscillator network has been proposed as an associative-memory network. The authors of~\cite{li2025associative} take $A(x)=\cos x$ and $\w_k=0$, but they consider, in contradistinction to our approach,  a symmetric coupling $\dot x_k=A'(x_k)\sum (a_{kj}+a_{jk})A(x_j)$. Such a system can be written as a gradient system and possesses attractive steady states only, a property diametrically opposed to conservative dynamics.

A similar form of the dynamics holds for a Kuramoto-Daido-type coupling: 
\begin{equation}
\dot{x}_k=\w_k+\sum_{j\neq k} \left(c_{kj} F_{even}(x_j-x_k)+d_{kj} F_{odd}(x_j-x_k)\right)\;,
\label{eq:10}
\end{equation}
where matrix $c_{kj}=c_{jk}$ is symmetric and matrix $d_{kj}=-d_{jk}$ is antisymmetric, and 
$F_{odd}(x)=-F_{odd}(-x)$, $F_{even}(x)=F_{even}(-x)$. The simplest case of odd coupling function is $F_{odd}(x)=\sin x$, which leads to a standard Kuramoto-type network with an antisymmetric coupling matrix. The simplest case of an even function is $F_{even}(x)=\cos x$, see a detailed discussion in Sec.~\ref{sec:rn}. Using the Fourier series of these coupling functions, one can represent the coupling by virtue of complex anti-Hermitian coupling matrices
\begin{equation}
\dot{x}_k=\w_k+\sum_{j} \text{Im}(q_{kj,s}\exp(is(x_j-x_k))),\quad q_{kj,s}+q_{jk,s}^*=0\;.
\label{eq:11}
\end{equation}
In particular, the Kuramoto-type model with conservative coupling that includes the first harmonics $s=1$ only, has the form
\begin{equation}
\dot{x}_k=\w_k+\sum_{j} \text{Im}[Q_{kj}\exp(i(x_j-x_k))],\quad Q_{kj}+Q_{jk}^*=0\;.
\label{eq:12}
\end{equation}

\subsection{Regular networks} 
\label{sec:rn}
It is instructive to consider the simplest cases of regular networks with Kuramoto-type coupling. As it follows from Eq.~\eqref{eq:10}, the case of conservative global coupling (all-ones coupling matrix) is possible if the coupling function is even, i.e., if it is represented by a cosine Fourier series. The simplest case is that of the first harmonic coupling: 
\begin{equation}
\dot{x}_k=\w_k+ \frac{\e}{N}\sum_{j=1}^N\cos(x_j-x_k) \;,\quad k=1,\ldots,N\;.
\label{eq:12-1}
\end{equation}
Such a coupling can be represented via the complex mean field
\begin{equation}
Z=Re^{i\Theta}=\frac{1}{N}\sum_j e^{i x_j}
\label{eq:12-2}
\end{equation}
as
\begin{equation}
\dot{x}_k=\w_k+\text{Im}\left(iH e^{-ix_k}\right),\quad H=\e Z\;.
\label{eq:12-3}
\end{equation}
For the analysis of this system, powerful analytical methods developed by Watanabe and Strogatz~\cite{Watanabe-Strogatz-93,Watanabe-Strogatz-94} and Ott and Antonsen~\cite{Ott-Antonsen-08}. Particulary simple equations for the order parameter $Z$ appear in the Ott-Antonsen theory, in the thermodynamic limit $N\to\infty$ for a Cauchy distribution of natural frequencies $g(\w)=\gamma/(\pi(\gamma^2+(\w-\w_0)^2))$:
\begin{equation}
\begin{gathered}
\dot Z=(i\w_0-\gamma)Z+\frac{\e}{2}(H-H^*Z^2)=\\-\gamma Z+i Z\left(\w_0+\frac{\e}{2}\left(1+|Z|^2\right)\right)\;.
\end{gathered}
\label{eq:12-4}
\end{equation}
This equation is valid on the Ott-Antonsen manifold, i.e., for a wrapped Cauchy distribution of the phases $x$.
One can see that the only dissipative term in this equation is related to the diversity of the oscillators' frequencies $\gamma$; if $\gamma=0$, Eq.~\ref{eq:12-4} describes a nonlinear conservative oscillator. We stress here that ``collective dissipation'', which reflects equilibration in the population of oscillators, does not contradict the conservative nature of the microscopic dynamics.

Another type of a regular network is a lattice with nearest-neighbor coupling. Assumung spatial homegineity, we set in  Eq.~\eqref{eq:10} $\w_k=0$, $c_{k,k+1}=c$, $d_{k,k+1}=d$, and obtain the following lattice equations
\begin{equation}
\begin{gathered}
\dot x_k=c(F_{even}(x_{k+1}-x_k)+F_{even}(x_{k-1}-x_k))+\\
d(F_{odd}(x_{k+1}-x_k)-F_{odd}(x_{k-1}-x_k))\;.
\end{gathered}
\label{eq:12-5}
\end{equation}
It is convenient to introduce the new variables $y_k=y_{k+1}-y_k$. Then Eq.~\eqref{eq:12-5} takes form
\begin{equation}
\dot y_k=F(y_{k+1})-F(y_{k-1}),\qquad F(y)=cF_{even}(y)+dF_{odd}(y)\;.
\label{eq:12-6}
\end{equation}
This lattice has been studied in Refs.~\cite{Rosenau-Pikovsky-05,Pikovsky-Rosenau-06,Ahnert-Pikovsky-08}. Its remarkable property is that it can be written in Hamiltonian form; see \cite{Pikovsky-Rosenau-06} for the derivation.

\subsection{Chaos at conservative coupling on networks}

Because of the conservation of the phase volume, attractors and repellers in the system are not possible; we generally expect to observe quasiperiodic and chaotic regimes, depending on initial conditions (if the system is not ergodic). We illustrate such a coexistence of regular and chaotic motions in a system of three Winfree-type coupled oscillators (Eq.~\eqref{eq:9}) with $A(x)=0.5+\sin(x)$, $\w_1=0.68,\;\w_2=1.1,\;\w_3=1.46$, and interaction matrix $a_{12}=0.45,\;a_{13}=0.225,\;a_{23}=-0.15$. Because for this system $x_3$ grows monotonically, we constructed a Poincar\'e map using the condition $x_3=0 \pmod{2\pi}$.
This map demonstrates domains of regular (quasiperiodic) and chaotic behavior, which is typical for Hamiltonian systems and area-preserving maps.

\begin{figure}[!htb]
\centering
\includegraphics[width=\columnwidth]{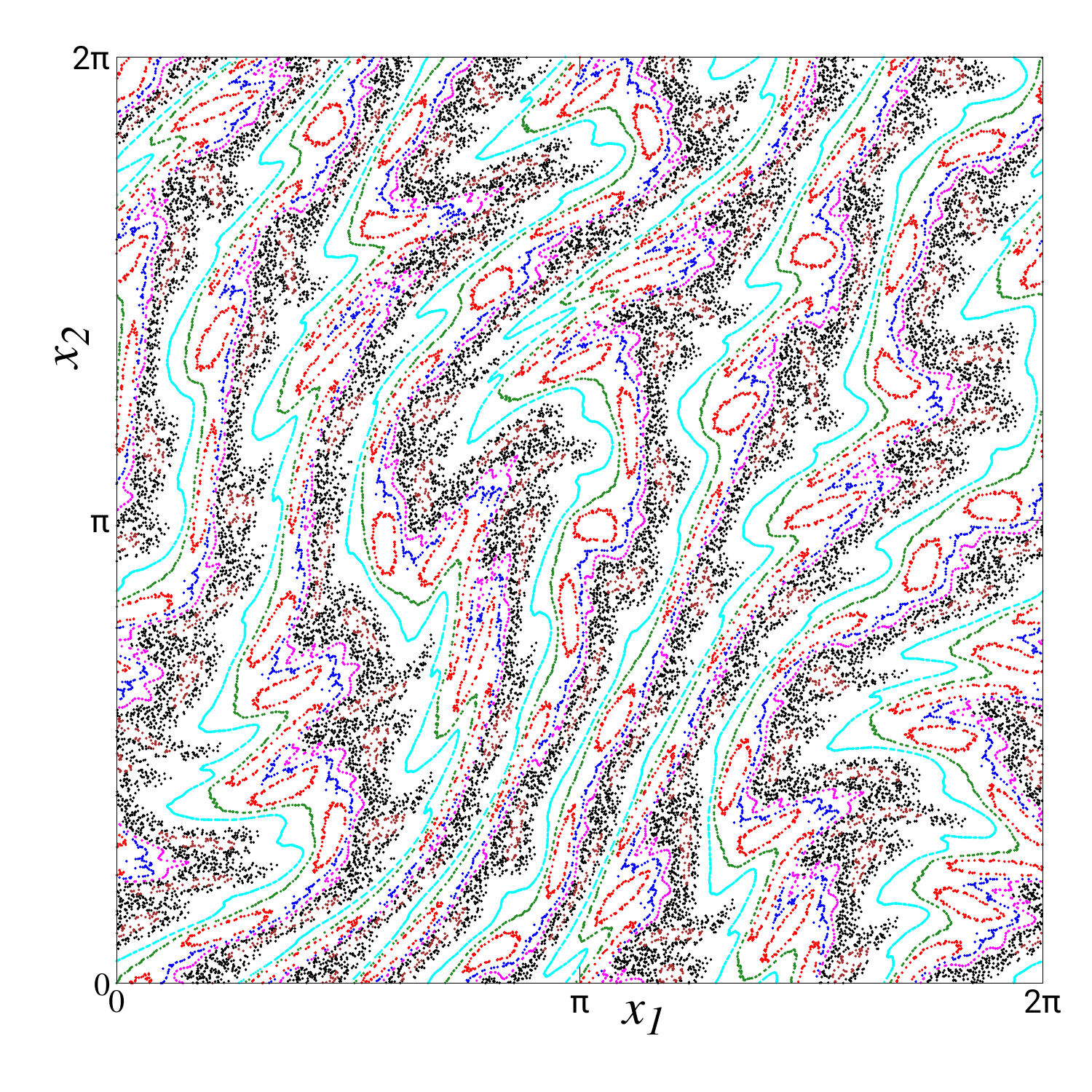}
\caption{Poincar\'e  map at the conservative Winfree-type coupling of three oscillators. Parameters are given in text. For some initial conditions, the trajectory forms a closed curve (torus in the phase space), while at other initial conditions, chaotic sets are produced.}
\label{fig:2}
\end{figure}

\begin{figure}[!htb]
\centering
\includegraphics[width=\columnwidth]{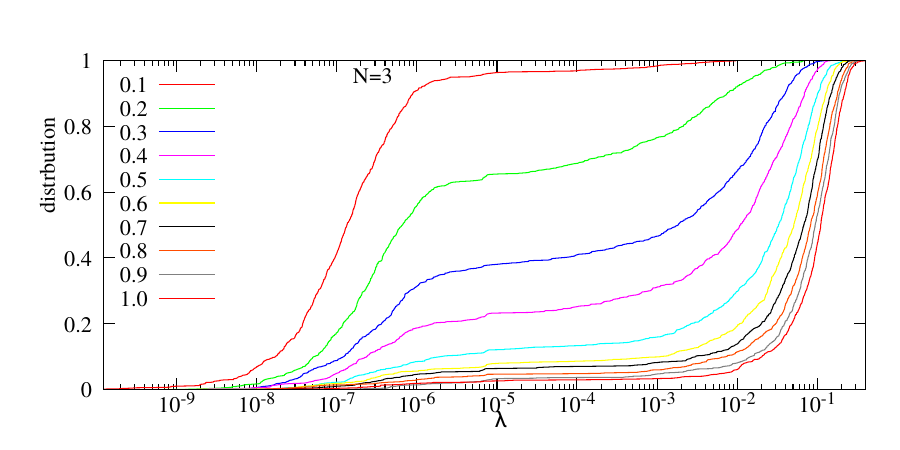}\\
\includegraphics[width=\columnwidth]{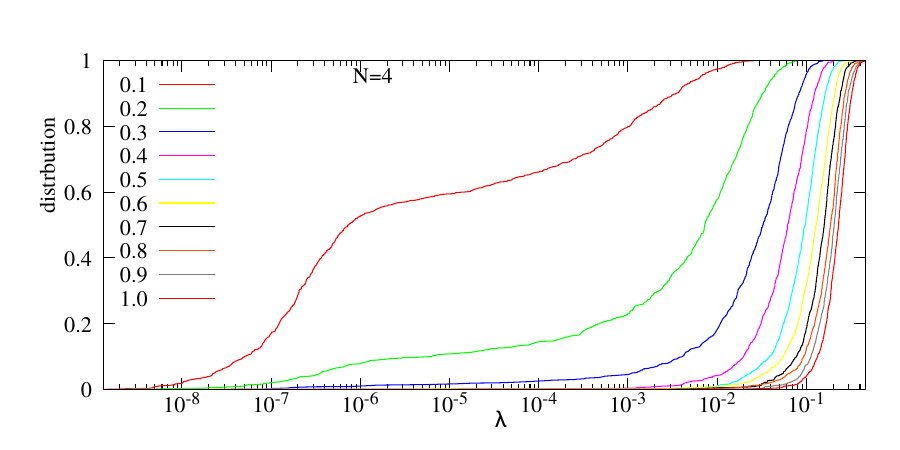}
\caption{Cumulative distributions of the largest Lyapunov exponent for $N=3,4$ and different coupling strengths $\e$ (lines of different colors), calculated from $1000$ independent runs as described in the text.}
\label{fig:3}
\end{figure}

One expects that, in the spirit of the KAM theory, at small coupling, quasiperiodic regimes are dominant, while chaos is typical for strong coupling. To check this, we calculated the largest Lyapunov exponent for networks \eqref{eq:9}  of three and four units, with   $A(x)=0.5+\sin(x)$, natural frequencies $\w$ sampled from the uniform distribution $0.5\leq \w\leq 1.5$, and the elements of the coupling matrix sampled from a Gaussian distribution $a_{kj}=\e \mathcal{N}(0,1)$. Here, the parameter $\e$ determines the strength of the coupling. We show the cumulative distributions of the largest Lyapunov exponent for different $\e$ and $N=3,4$ in Fig.~\ref{fig:3}.

Because the numerically calculated largest Lyapunov exponent is never zero, one interprets small values (in our case $<10^{5}$) as an indication of the quasiperiodic dynamics. Correspondingly, larger values of the Lyapunov exponent indicate chaos. One can see that for $N=3$, quasiperiodicity dominates for $\e=0.1$; for $\e=0.2$, it occurs (according to the adopted criterion) with probability $0.65$; and for $\e=0.3$, it occurs with probability $0.38$. For $N=4$, already for $\e=0.1$ the probability of quasiperiodicity is $0.6$, and it drops to $0.1$ for $\e=0.2$. For stronger couplings and $N=4$, practically in all cases, chaos is observed. 

The qualitative properties of chaos at conservative coupling via pair-Hamiltonians are similar to those in genuine Hamiltonian systems. However, there is an important difference regarding the Lyapunov exponents. In Hamiltonian systems, due to the symplectic structure, the Lyapunov exponents come in pairs 
$(\lambda,-\lambda)$~\cite{Pikovsky-Politi-16}. In the conservative phase dynamics system as introduced above, there is no symplectic structure; only the conservation of the phase volume is ensured. This property implies that the sum of all Lyapunov exponents vanishes $\sum_{i=1}^N\lambda_i=0$, but it does not imply that there is pairwise symmetry of positive and negative exponents.

\begin{figure}[!htb]
\centering
\includegraphics[width=\columnwidth]{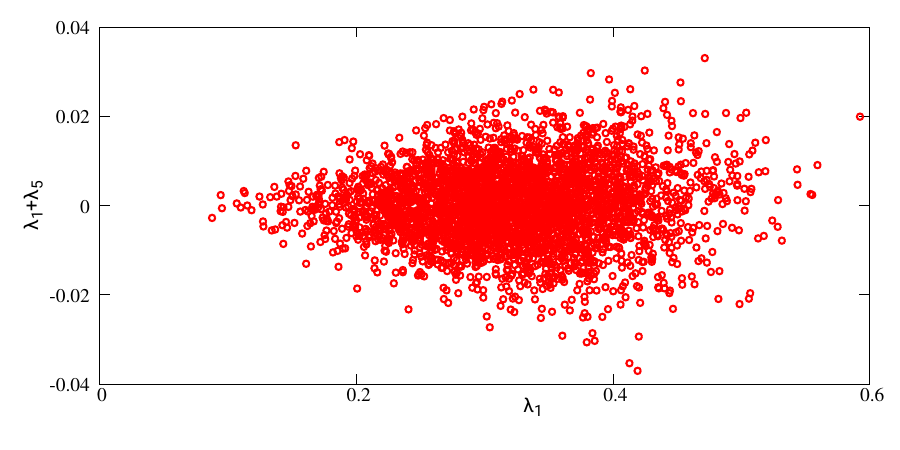}
\caption{The sums of the largest and of the smallest Lyapunov exponents vs the largest exponent for conservative networks of 5 phase oscillators.}
\label{fig:4}
\end{figure}

We illustrate the absence of the pairwise symmetry of Lyapunov exponents in Fig.~\ref{fig:4}. Here, the Lyapunov exponents are reported for networks of $N=5$ oscillators with Winfree-type conservative coupling (the same parameters as used in Fig.~\ref{fig:3}). From five Lyapunov exponents, the two first $\lambda_1,\lambda_2$ are always positive; the two last $\lambda_4,\lambda_5$ are always negative, and $\lambda_3$ is close to zero (of order $10^{-6}$). In the figure, we plot $\lambda_1+\lambda_5$ vs $\lambda_1$, to show that $\lambda_1+\lambda_5$ is not close to zero. The sum of all Lyapunov exponents in these calculations is typically in the range $\left|\sum_i\lambda_i\right|<10^{-7}$.

\begin{figure}[!htb]
\centering
\includegraphics[width=\columnwidth]{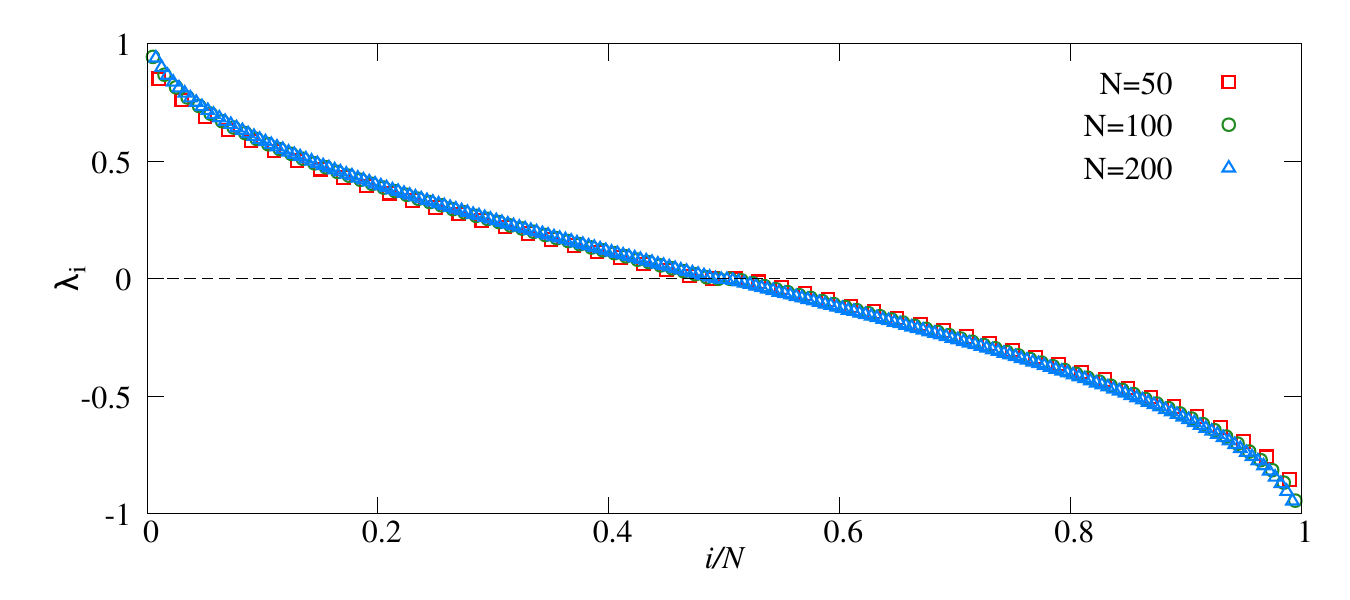}
\caption{Spectra of Lyapunov exponents for several values of $N$, for random networks with Kuramoto-type coupling~\eqref{eq:12}. The spectrum appears nearly anti-symmetric with respect to the midpoint $i=N/2$.}
\label{fig:5}
\end{figure}

We note that for large networks, the symmetry between positive and negative Lyapunov exponents is violated much less than for small networks, possibly due to self-averaging of the exponents. We illustrate this in Fig.~\ref{fig:5}, where we show full spectra of Lyapunov exponents for the networks with random Kuramoto-type couplings \eqref{eq:12}
where $\w_k=0$ and real and imaginary parts of the coupling constants are independent Gaussian random numbers $Q_{kj}=N^{-1/2}(\mathcal{N}(0,1)+i\mathcal{N}(0,1))$. Visually, the spectra appear pretty symmetric, although detailed analysis reveals that the values of $|\lambda_i+\lambda_{N-i}|$ are of order $10^{-3}$.

\section{Conservative many-body coupling}
\subsection{General consideration}
Recently, many-body coupling of phase oscillators attracted large attention~\cite{Pikovsky-Rosenblum-22,bick2023higher}. Here, we show how the concept of pair-Hamiltonians can be generalized to many-body interactions. We restrict ourselves to triplet couplings only; extension to quadruplets, etc., is straightforward. For three phase oscillators with phases $x,y,z$ we introduce triplet-Hamiltonian $p(x,y,z)$ as an arbitrary $2\pi$-periodic function of the arguments, and write the dynamics as
\begin{equation}
\begin{aligned}
\dot x&=\w_x+a_x\partial_{yz} p(x,y,z)\;,\\
\dot y&=\w_y+a_y\partial_{xz} p(x,y,z)\;,\\
\dot z&=\w_z+a_z\partial_{xy} p(x,y,z)\;,\\
&a_x+a_y+a_z=0\;.
\end{aligned}
\label{eq:13}
\end{equation}
It is easy to check that the phase volume in this system is conserved
\[
\partial_x\dot x+\partial_y\dot y+\partial_z\dot z=(a_x+a_y+a_z)\partial_{xyz} p(x,y,z)=0\;.
\]
Conservation of the phase volume in a large network is ensured if all pairwise interactions are of type~\eqref{eq:4},
all triplet interactions are of type \eqref{eq:13}, etc.

\subsection{Regular hypernetwork with triplet coupling}

Here, we give a simple example of a network with triplet coupling. Following~\cite{Komarov-Pikovsky-13}, we consider three communities of oscillators $x_k,\;y_k,\;z_k$, natural frequencies of which are concentrated around $\w_x,\;\w_y,\;\w_z$, respectively. If these frequencies fulfil a resonance condition $\w_z=\w_x+\w_y$, then the leading terms in the interactions between the oscillators in different communities are triplets:
\begin{equation}
\begin{aligned}
\dot z_k&=\w_{z,k}+\sum_{j,m} (a_z\sin(x_j+y_m-z_k)+\\
&b_z\cos(x_j+y_m-z_k))\;,\\
\dot y_k&=\w_{y,k}+\sum_{j,m} (a_y\sin(z_j-x_m-y_k)+\\ 
&b_y\cos(z_j-x_m-y_k))\;,\\
\dot x_k&=\w_{x,k}+\sum_{j,m} (a_x\sin(z_j-y_m-x_k)+\\
&b_x\cos(z_j-y_m-x_k))\;.
\end{aligned}
\label{eq:13-1}
\end{equation}
Here we assume that the coupling is global (i.e., the coupling constants $ a$ and $ b$ do not carry indices) and acts via the first harmonics. Calculating the divergence of the phase volume for the triplet coupling network \eqref{eq:13-1}, we obtain
\begin{equation}
\begin{gathered}
\sum_k\left(\frac{\partial\dot z_k}{\partial z_k}+\frac{\partial\dot y_k}{\partial y_k}+\frac{\partial\dot x_k}{\partial x_k}\right)=\\
\sum_{l,m,k}\big[(a_x+a_y+a_z)\cos(z_k-x_l-y_m)+\\
(b_z-b_y-b_x)\sin(x_k+y_l-z_m)\big]\;.
\end{gathered}
\label{eq:13-2}
\end{equation}
This divergence vanishes if $a_x+a_y+a_z=b_z-b_x-b_y=0$. These are the conditions for conservative triplet interactions among three communities, whose frequencies obey the resonant condition above.

\subsection{Regular network with quadruplet coupling}

A general physical mechanism resulting in effective many-body interaction of oscillators is coupling via a nonlinear function of the mean field~\cite{Pikovsky-Rosenblum-22}. In the Kuramoto model Eq.~\eqref{eq:12-3}, the complex force term $H$ acting on each oscillator is a linear function of the complex mean field $H=\e Z$. Let us use a nonlinear relation $H=\e |Z|^2Z$ instead. Then, instead of two-body coupling \eqref{eq:12-1} we obtain a quadruplet coupling
\begin{equation}
\dot{x}_k=\w_k+ \frac{\e}{N^3}\sum_{j,l,m=1}^N\cos(x_j+x_l-x_m-x_k) \;,\quad k=1,\ldots,N\;.
\label{eq:13-3}
\end{equation}
An elementary calculation shows that the phase volume in hypernetwork \eqref{eq:13-3} is conserved.

\section{Conservative coupling of Stuart-Landau oscillators}

Above, we considered phase equations. Here, we discuss phase reduction for Stuart-Landau (SL) oscillators and show under which conditions the resulting coupling in the phase equations is conservative. 

We write a network of coupled SL oscillators, described by their complex amplitudes $a_k$, as
\begin{equation}
\dot a_k=i\omega_k a_k+\mu_k a_k-\gamma_k a_k|a_k|^2+\sum_m(D_{km}+i C_{km}) a_m\;.
\end{equation}
We introduce the real amplitude and the phase via $a_k=R_ke^{i\theta_k}$ and obtain
\begin{align}
\dot R_k&=\mu_k R_k-\gamma_k R_k^3 +\sum_m R_m\big(C_{km} \sin(\theta_m-\theta_k)+\nonumber\\
&D_{km}\cos(\theta_m-\theta_k)\big)\;, \label{eq:15-1}\\
\dot \theta_k&=\omega_k+\sum_m \frac{R_m}{R_k}
\big(C_{km}\cos(\theta_m-\theta_k)+\nonumber\\
&D_{km}\sin(\theta_m-\theta_k)\big)\;. \label{eq:15-2}
\end{align}

We assume that the coupling terms $C,D$ are small, thus from \eqref{eq:15-1} $R_k\approx(\mu_k/\gamma_k)^{1/2}$.
Substituting this into the phase equations \eqref{eq:15-2}, we obtain in the leading order
\begin{equation}
\begin{gathered}
\dot \theta_k=\omega_k+\sum_m (\mu_m/\gamma_m)^{1/2}(\mu_k/\gamma_k)^{-1/2}
\big(C_{km}\cos(\theta_m-\theta_k)+\\
D_{km}\sin(\theta_m-\theta_k)\big)
\end{gathered}
\label{eq:16}
\end{equation}
This system belongs to the class of Kuramoto-type networks~\eqref{eq:12} and is conservative if the coupling coefficients possess the following symmetry:
\begin{equation}
\begin{gathered}
C_{km}(\mu_m/\gamma_m)=
C_{mk}(\mu_k/\gamma_k)\;,\\
D_{km}(\mu_m/\gamma_m)=
-D_{mk}(\mu_k/\gamma_k)\;.
\end{gathered}
\label{eq:17}
\end{equation}
If all the oscillators have identical amplitudes $R_1=\ldots=R_N$, then condition \eqref{eq:17}
reduces to anti-Hermicity of the original complex coupling matrix: $C_{mk}=C_{km}$, $D_{mk}=-D_{km}$.

We stress here that the conservative nature of coupling is ensured only in the first-order (in the powers of the coupling strength) phase reduction; in higher orders, this property is violated~\cite{Pikovsky-Rosenau-06,kuznetsov2016oscillations}.

\section{Conclusion}
In this paper, we have introduced a general class of phase equations with conservative dynamics that conserves the phase volume. Each pairwise coupling term is represented via derivatives of a function of two phases; we call this function pair-Hamiltonian. There is no global Hamiltonian structure of the equations in the usual sense, although in some particular cases, such a global Hamiltonian description might be possible~\cite{Pikovsky-Rosenau-06}. Notably, pair-Hamiltonians can contain explicit time dependence. 

We also discussed simplified versions of pair-Hamiltonians, describing Winfree-type and Kuramoto-Daido-type couplings. In the case of a network with pure Kuramoto-type coupling, the complex coupling matrix (real and imaginary parts of which describe sin and cos coupling terms, respectively) must be anti-Hermitian to ensure conservativity. The dynamics of conservative networks is typically chaotic if the number of units is large. Remarkably, in contradistinction to standard Hamiltonian chaos, here there is no pairwise symmetry between the Lyapunov exponents; only their sum vanishes. Furthermore, we discussed the generalization of the concept to many-body (triplet) interactions. We also demonstrated that the first-order (in coupling strength) phase reduction of a network of coupled Stuart-Landau oscillators can yield a conservative system if the coupling coefficients satisfy certain conditions.  From the viewpoint of applications, conservative phase dynamics appears to be suitable for an experimental implementation in electronic circuits~\cite{minati25three}, with possible use for chaos-based secure communication, cryptography, and pseudo-random number generation.

%
%

\acknowledgements{The author thanks O. Burylko, S. Flach, and Ph. Rosenau for useful discussions.}







%

\end{document}